# High power Figure-of-Merit, 10.6-kV AlGaN/GaN lateral Schottky barrier diode with single channel and sub-100-μm anode-to-cathode spacing

Ru Xu[1], Peng Chen[1,✉], Jing Zhou[1], Yimeng Li[1], Yuyin Li, Tinggang Zhu[2], Kai Cheng[3], Dunjun Chen[1], Zili Xie[1], Jiandong Ye, Bin Liu[1], Xiangqian Xiu, Ping Han, Yi Shi, Rong Zhang[1,✉], Youdou Zheng[1,✉]

[1]The Key Laboratory of Advanced Photonic and Electronic Materials, School of Electronic Science and Engineering, Nanjing University, Nanjing, China.
[2]Corenergy Semiconductor Incorporation, Suzhou, China.
[3]Enkris Semiconductor Inc.NW-20, Nanopolis, Suzhou, China.
✉e-mail:: pchen@nju.edu.cn; rzhang@nju.edu.cn; ydzheng@nju.edu.cn.

**GaN-based lateral Schottky diodes (SBDs) have attracted great attention for high-power applications due to its combined high electron mobility and large critical breakdown field. However, the breakdown voltage (*BV*) of the SBDs are far from exploiting the material advantages of GaN at present, limiting the desire to use GaN for ultra-high voltage (UHV) applications. Then, a golden question is whether the excellent properties of GaN-based materials can be practically used in the UHV field? Here we demonstrate UHV AlGaN/GaN SBDs on sapphire with a *BV* of 10.6 kV, a specific on-resistance of 25.8 mΩ·cm$^2$, yielding a power figure of merit of more than 3.8 GW/cm$^2$. These devices are designed with single channel and 85-μm anode-to-cathode spacing, without other additional electric field management, demonstrating its great potential for the UHV application in power electronics.**

Group-III nitrides represented by GaN have been considered to be one of the most important semiconductor materials after silicon. The two-dimensional electron gas (2-DEG) generated in the AlGaN/GaN heterostructure has the characteristics of high concentration ($>10^{13}/cm^2$) and high electron mobility ($>2000$ $cm^2/v·s$), making it an excellent choice for power electronic devices. The lateral Schottky barrier diode (SBD) based on the AlGaN/GaN structure has received extensive attention[1-25]. Furthermore, GaN has a critical breakdown electric field (*EF*) of up to 3.4 MV/cm[26], which is much higher than other semiconductor materials, such as Si (0.3 MV/cm), GaAs o (0.4 MV/cm) and SiC (2.0 MV/cm). Based on the above-mentioned material advantages, currently AlGaN/GaN-based electronic devices can be widely used in high-power and high-frequency electronics fields of medium and high voltage range. Then, a golden question (GQ) was raised, **that is whether the excellent properties of GaN-based materials can be practically used in the field of ultra-high voltage (UHV)?** After all, theoretically, the breakdown voltage (*BV*) can reach 10 kV based on the AlGN/GaN lateral SBD with a 60-μm electrode spacing, which will be shown later.

UHV (>10 kV) semiconductor power devices are key component for the application in utility grid, pulsed power, industrial motor drives, very high voltage switches for high-voltage power transmission, power substations and high-temperature environment. Previous reports show that by increasing the distance between the corresponding electrodes, the breakdown voltage of GaN-based power devices can be linearly increased [11,27-29]. It is an obvious way to increase the *BV* by increasing the anode-to-cathode spacing ($L_{AC}$) for lateral SBDs. However, successful applications depend on the UHV device that needs to have multiple characteristics at the same time, such as low forward voltage, high switching frequency, low on-resistance and easy well series/parallel connection, etc.. However, some characteristics are mutually restrictive in the device design, especially for high *BV* and low on-resistance, they increase synchronously in many cases. Thus, people cannot simply increase $L_{AC}$ to gain UHV performance. Based on these characteristics and the importance of the application of 10-kV devices, we propose **a baseline, that is to achieve a *BV* of over 10 kV with a $L_{AC}$ of less than 100 μm**.

Lots of high-performance AlGaN/GaN lateral SBD devices have been reported [1-22], such as SBDs on Si with breakdown voltages up to 3.4 kV, and SBDs on sapphire with breakdown voltages more than 9 kV [11,29,30]. However, the GQ proposed above still has not been answered. For an example, in 2008, Ishida et al. reported a 9.3-kV GaN diode, as well as the effects of single-channel and dual-channel [29]. However, the $L_{AC}$ of their devices is as long as 160 μm, resulting in an specific on-resistance ($R_{ON,SP}$) as high as 176 mΩ·cm$^2$, which still makes it doubtful whether the GaN material is suitable for the UHV field. After all, SiC junction barrier Schottky diodes and MOSFET up to 10 kV have been developed [31,32]. Recently, the performance of GaN-based SBD devices has been significantly improved. Just in last month, Xiao et al. reported a GaN-based lateral SBD device with a $BV$ up to 10 kV with a greatly improved $R_{ON,SP}$ [33]. However, $L_{AC}$ of their devices still exceeds 100 μm to 123 μm, resulting in device $R_{ON,SP}$ as high as 39 mΩ·cm$^2$. At the same time, some research results indicate that $L_{AC}$ is not the only factor limiting $BV$, other factors such as the influence from the substrate. The research of GaN-based UHV devices on Si substrates seems to have encountered difficulties at present. Fundamentally, how different substrates affect the $BV$ is also a question that needs to be answered. But, whether it is possible for actual GaN-based power devices to approach the theoretical limit in the UHV field has come to the time for an urgent answer. Whatever, driven by these outstanding research work, people continue to explore ways to push the performance of GaN SBD devices to their theoretical limits. It is clear that the material quality and device design are key factors in determining the device performance, especially for the $BV$ and $R_{ON,SP}$. In this work, we fabricated AlGaN/GaN lateral SBDs on high-quality GaN-based materials grown on sapphire substrates, and demonstrate a high power FOM of 3.8 GW/cm$^2$, 10.6-kV AlGaN/GaN lateral Schottcky barrier diode with a $L_{AC}$ of 85 μm. We also studied the inherent problems of GaN SBD on Si substrate.

## Device design and fabrication

The samples were grown by metal organic chemical vapor deposition (MOCVD) on 2-inch c-plane sapphires and 6-inch p-Si substrates. For the devices on sapphire, the device structure consists of nucleation layer, a 3-μm C-doped GaN buffer layer, a 100-nm i-GaN channel layer, a 1-nm AlN spacer, a 20-nm $Al_{0.25}Ga_{0.75}N$ barrier layer, a 2-nm GaN cap layer and a 50-nm in-situ $SiN_x$. For the devices on Si, the device structures consist of a (Al, Ga) N buffer layer with thicknesses of 5 μm, a 200-nm i-GaN channel layer, an 1-nm AlN spacer, a 25-nm $Al_{0.25}Ga_{0.75}N$ barrier layer, a 2-nm GaN cap layer and a 50-nm in-situ $SiN_x$ passivation layer.

It is well known that the anode field plate (AFP) can effectively modify the *EF* distribution and then increase the *BV*. In our previous work[34-36], prior to the fabrication of the SBDs on Si, we also used Silvaco-TCAD to simulate and optimize the AFP parameters. The simulated *BV* is highly depended on the AFP length and SiN$_x$ passivation layer thickness. When the optimized AFP structure is added to the device, the surface *EF* value is greatly weakened by the decrease of about 38% and the *BV* can be boosted about 30% compared to the device structure without the AFP. Based on these optimized parameters, we fabricated the SBDs on Si.

The detailed fabrication process is shown in the section of Method, and some process parameters and the main results have been reported in our previous work [34-37].

For the fabrication of the SBDs on sapphire, we have different ideas evolved from our previous works on Si. First, we have known that the *EF* in the device is still far below the critical value of 3.4 MV/cm at the breakdown point, even with the optimized AFP structure. Secondly, we have known that the anode fabrication process play an very important role in the characteristics of the *BV* and $R_{ON,SP}$, and we have developed a good fabrication process in our previous works. Thirdly, due to the more mature epitaxial technology and better lattice/thermal matching of GaN on sapphire than those on Si, the material quality of the GaN grown on sapphire is deemed to be better than that on Si. Therefore, we decided to fabricated the AlGaN/GaN lateral SBDs WITH single channel and WITHOUT the AFP. Another purpose of not using AFP is better to test the intrinsic potential of GaN materials, rather than relying too much on additional structures to do electric field management. The detailed fabrication process is shown in the section of Method. Fig. 1a shows the schematic cross section of the devices with circular anode on sapphire. Fig.1b shows the photograph of a device with a $L_{AC}$ of 85 μm on sapphire, and the anode diameter is 180 μm, the total area of the SBD is 420 × 420 μm$^2$.

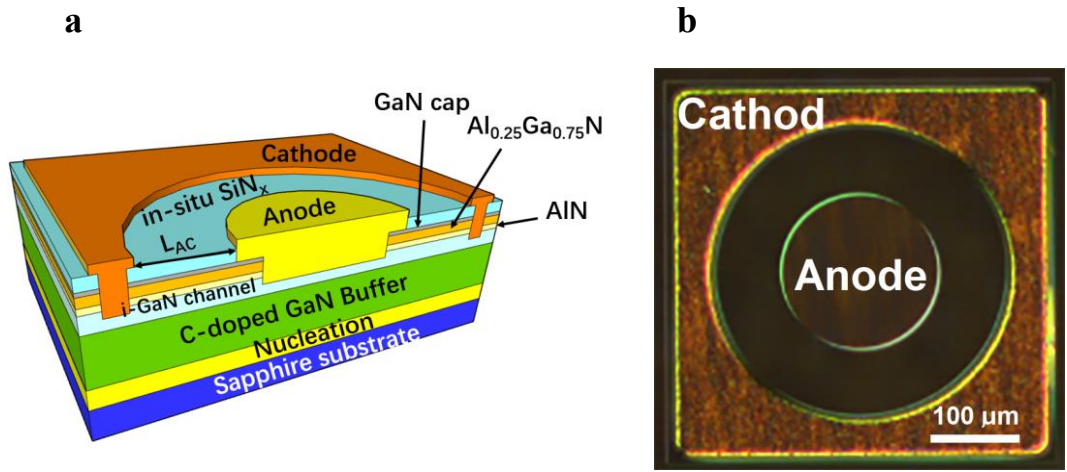

**Fig. 1 | Device structure. a,** Schematic cross section view of the AlGaN/GaN lateral SBDs on sapphire. **b,** photograph of top view of the fabricated SBD with 85-μm $L_{AC}$ and 180-μm-diameter anode.

## Room-temperature forward and reverse *I-V* characteristics

At room temperature, we measured the forward and reverse *I-V* characteristics of the devices. Fig.2 presents the results from the SBDs on sapphire with 85-μm $L_{AC}$. In order to obtain typical device performance, 40 devices have been tested to show the overall forward *I-V* characteristics, as shown in Fig.2a. All measured results distribute within the red curves and the blue curves, and this relatively small device-to-device fluctuation shows relatively stable device performance. The average turn-on voltage ($V_{on}$) is 0.72 V at the forward current of 1 mA/mm, and the average $R_{ON,SP}$ is 25.8 mΩ·cm² by considering a 1.5-μm transfer length of ohmic contact and a 1.5-μm extension length of the Schottky contact.

In order to investigate the relationship between the *BV* and $L_{AC}$, we fabricated AlGaN/GaN lateral SBDs on sapphire substrates with $L_{AC}$ values ranging from 5 μm to 85 μm. Fig.2b shows the reverse *I-V* characteristics of the SBDs on sapphire with the $L_{AC}$ from 5 μm to 85 μm. It can be seen that the *BV* roughly increases in proportion to the $L_{AC}$. Eventually, the device with 85-μm $L_{AC}$ gives the *BV* more than 10 kV to 10.6 kV, and the leakage current is 0.1 μA/mm at 90% of the *BV*. Combining with the $R_{ON,SP}$ of 25.8 mΩ·cm², a calculated P-FOM ($BV^2/R_{ON,SP}$) of 3.8 GW/cm² is achieved in the 85-μm-$L_{AC}$ SBD device.

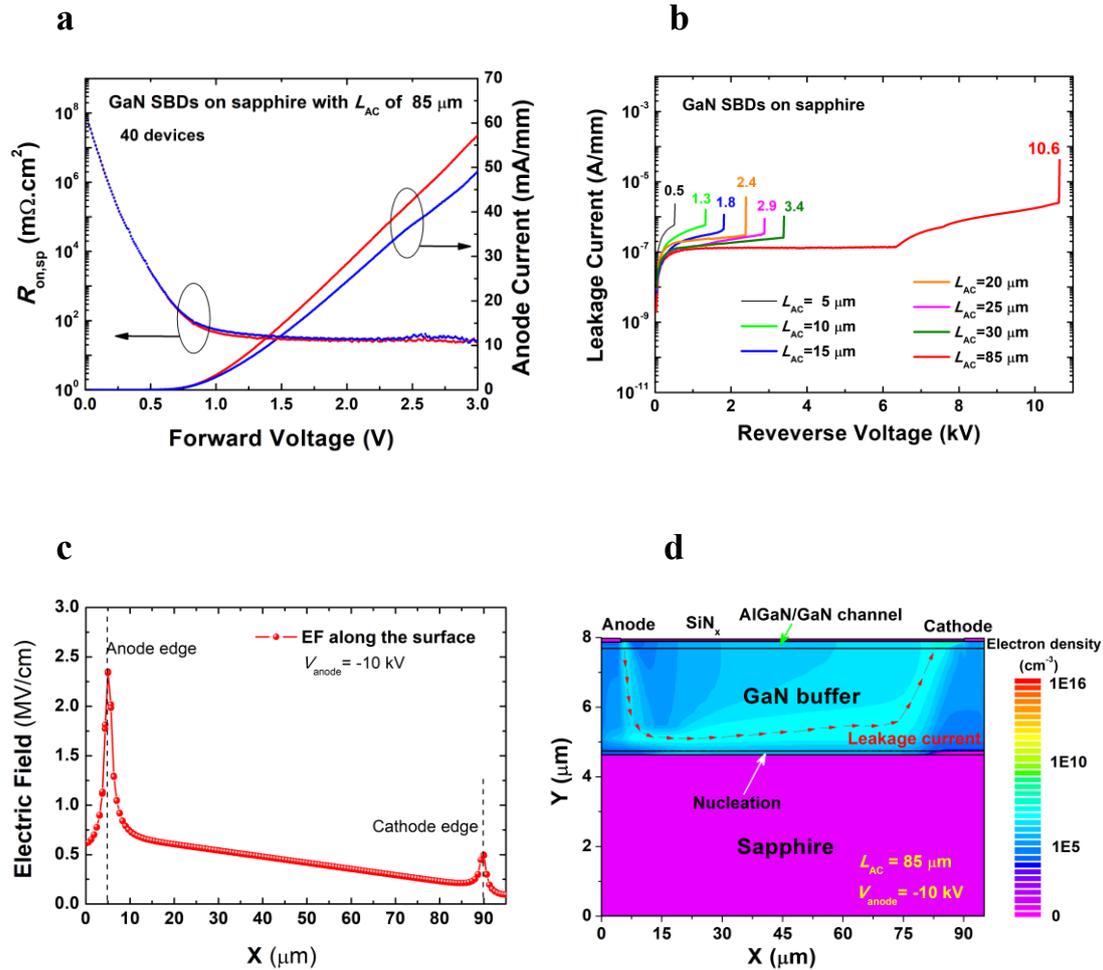

**Fig. 2 | Tested and simulated results of AlGaN/GaN SBDs on sapphire substrates. a** Forward *I-V* characteristics of 40 GaN-on-sapphire lateral SBDs with the electrode spacing of 85 μm. **b** Reverse *I-V* characteristics of the lateral SBDs with the electrode spacing of 5 to 85 μm. **c** Electric filed distribution along the AlGaN/GaN surface of the 85-μm-$L_{AC}$ SBD device at -10 kV. **d** Electron distribution of the 85-μm-$L_{AC}$ SBD device at -10 kV.

It has been known that in an ideal SBD device, when the reverse voltage is high enough, the *EF* around the anode edge surface will reach the critical *EF* intensity of GaN (3.4 MV/cm), and then the breakdown will occur with a sudden bump of leakage current. In order to figure out the actual *EF* situation, Silvaco-TCAD simulations were carried out based on the 10.6-kV device. As shown in Fig. 2c, the *EF* distribution on the surface was obtained at -10 kV. The peak value of the *EF* at the anode edge is about 2.4 MV/cm, lower than the critical *EF* intensity of GaN, indicating that the sudden bump of leakage current occurred elsewhere. This also indicates that the *BV* of our SBDs should still have room for further improvement.

Normally, electrons should be confined in the AlGaN/GaN channel under reverse bias and only move in the channel [38]. However, as shown in Fig.2b, we noticed an abnormal increase in leakage current after -6 kV. This abnormal increase has been reported to be closely related to the leakage in the GaN buffer, i.e. related to the trap ionization in the GaN buffer under high $EF$ [24, 25]. In this work, the buffer is the 3-μm GaN layer with a C doping concentration of $5\times10^{19}$ cm$^{-3}$. During the simulation, the C doping was set as deep acceptor-like traps with an energy level of 0.6 eV above the valance band [39]. As a typical GaN background, a donor-like trap density of $1\times10^{16}$ cm$^{-3}$ with an energy level of 1.2 eV below the conduction band was also set into the buffer [40]. Through the simulation, we find that the deep acceptor-like traps will be fully ionized at -10 kV, leading to an additional current channel from the anode to the cathode in the GaN buffer, as shown in Fig.2d. This effect contributes to the leakage current increase after -6 kV. This means that the acceptor-like traps introduced by C in the buffer will be no longer sufficient to confine electrons at extremely high voltages. Obviously, this additional leakage path will no doubt damage the breakdown characteristics of the device. As a countermeasure, the GaN buffers must be carefully designed to be able to withstand high $EF$ without generating additional leakage channels.

Theoretically, the $BV$ here is formulated with the similar model used in Ref. 6,

$$BV = \int_0^{L_{AC}} E(x)dx = \int_0^{L_{AC}} \frac{qN}{\varepsilon_0 \varepsilon_s}(L_{AC} - x)dx \qquad (1)$$

Where $E(x)$ is the surface $EF$ of the SBD. According to this formula, the $BV$ should increase linearly with the increase of the electrode spacing $L_{AC}$. The measured $BV$ and theoretical $BV$ for the GaN SBDs are shown in Fig.3a. The theoretical $BV$ line gives a slope of 170 V/μm, then the $BV$ of 10 kV can be predicted at the $L_{AC}$ of 60 μm.

To further verify the impact of the buffer on the breakdown voltage, we also fabricated SBDs on the p-Si substrates with $L_{AC}$ ranging from 5 μm to 85 μm, respectively. The measured $BV$ results of the SBDs on the p-Si substrates are also shown in Fig.3a. Fig.3b shows the reverse $I$-$V$ characteristics of the SBDs on the p-Si substrates with the $L_{AC}$ from 5 μm to 85 μm.

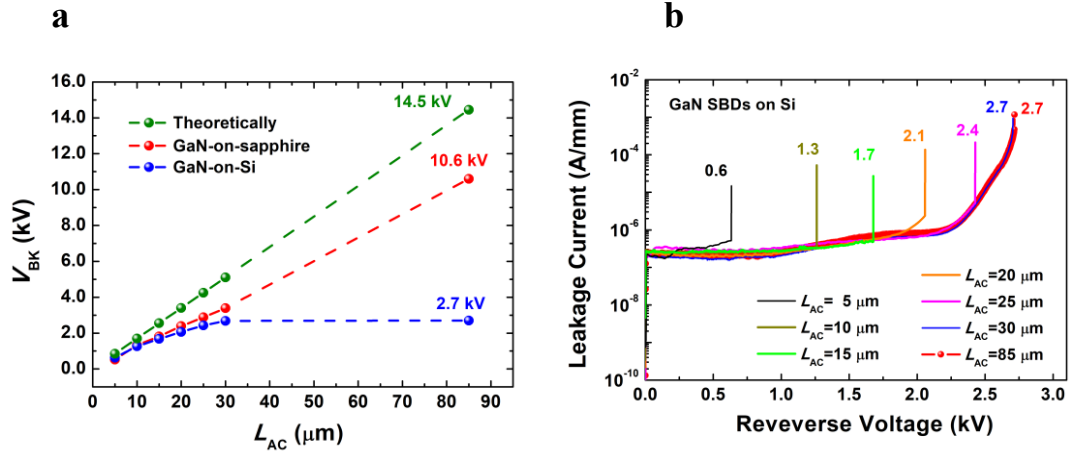

**Fig. 3 | Measured *BV* of SBDs on different substrates. a** The measured *BV* and theoretical *BV* for the GaN SBDs on sapphire and Si with different $L_{AC}$. **b** Reverse *I-V* characteristics of the SBDs on Si substrates with different $L_{AC}$ from 5 to 85 μm.

From Fig.3a, it can be seen that the *BV* values of the SBDs on sapphire increase quasi-linearly with the increase of $L_{AC}$ similar to the trend of theoretical *BV* with $L_{AC}$, which gives an experimental slope of 125 V/μm. For SBDs on Si, the *BV* also increases quasi-linearly at the $L_{AC}$ below 30 μm, and the fit yields a slope of 90 V/μm. These results present a sign that the GaN buffer quality on Si is lower than the one on sapphire, although the thickness of the GaN buffer (5 μm) on Si is larger than that (3 μm) on sapphire. After $L_{AC}$ > 30 μm, the *BV* of the SBDs on Si appears to be saturated at 2.7 kV. At the same time, the leakage currents of the SBDs with larger $L_{AC}$ (> 25 μm) exhibits two current increase phases, the first one is started at 1.2 kV, another one is started at 2.3 kV till to the breakdown, as shown in Fig.3b. The reason for the first current increase phase is similar to SBDs on sapphire, i.e. caused by ionized traps in the buffer [25]. However, the breakdown at this time seems to be more closely related to the second current increase phase, which causes the *BV* to be saturated at 2.7 kV.

## Device simulation studies

The huge differences between SBDs on sapphire and SBDs on Si can be seen, which suggests that other mechanisms may be involved. In order to reveal the deeper reasons, we did further simulations to obtain the essential difference between the two structures under high reverse voltage.

In the simulation process, we set the anode voltage of the two device structures to -2.5 kV, which is exactly in the second current increase phase of the device on Si. The simulated electron density of the SBDs on Sapphire and Si is shown in Fig.4.

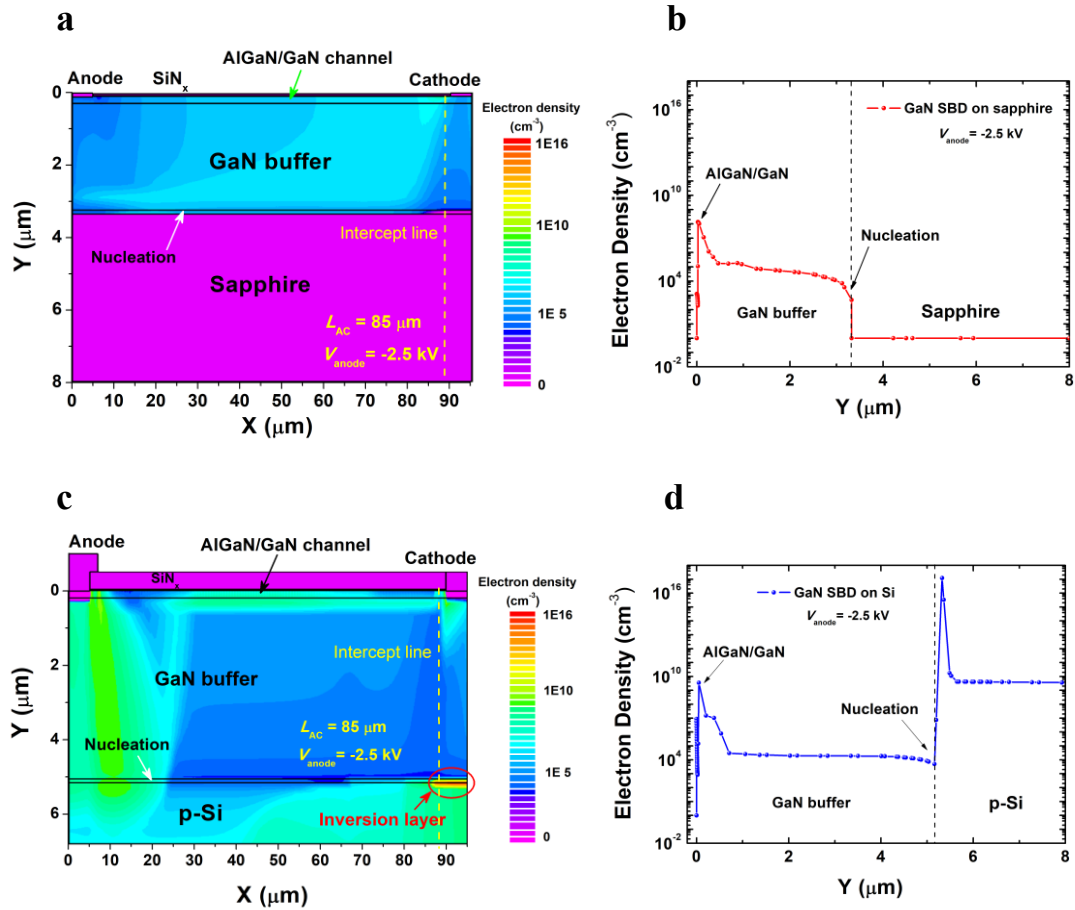

**Fig. 4 | Simulated results of AlGaN/GaN SBDs on Si and sapphire substrates. a** Simulated electron density of the SBDs on Sapphire and **c** sapphire substrates at -2.5 kV. Extracted electron concentration distribution within the same cathode edge for **b** GaN-on-Sapphire and **d** GaN-on-Si SBDs.

From Fig.4a, the result seems to be expected, i.e. the highest electron density around the AlGaN/GaN interface. Fig.4b shows the electron density profile extracted along the intercept line in Fig.4a. However, for the device on Si, high-density electrons were generated just below the interface between the GaN buffer and Si, as shown in Fig.4c. Since the Si substrate is p type, this is an inversion layer beneath the cathode. The electron density distribution inside the device was further extracted along the intercept line in Fig.4c, as shown in Fig.4d. The highly concentrated electron layer near the interface can reach about $10^{17}$ cm$^{-3}$. Obviously, this high-concentration electron layer will become another leakage channel through the Si substrate. The simulation results confirm that there is a parasitic conductive channel through the Si substrate under high reverse voltage. The reason is obvious too. Sapphire is an insulator and is not easily affected by electric fields. But Si is a semiconductor with narrower bandgap than the GaN buffer and the nucleation layer (AlN), electrons generated from any sources under high *EF* will accumulate in Si [24,41]. In order to minimize the influence of the Si substrate, a thicker GaN buffer is necessary. Just like our previous work [34-36], the *BV* can be improved from 2.7 kV to 3.4 kV when the GaN buffer thickness changes from 5 μm to 7 μm with similar device processing. At present, the leakage introduced by the Si substrate may become a serious problem, limiting the development of UHV AlGaN/GaN SBD on Si.

Based on the above discussion, benefiting from the higher material quality and the elimination of the substrate leakage, even without the assistance from AFP and additional channels, our GaN SBD devices on sapphire by proper processing can achieve the *BV* up to 10.6 kV with 85-μm $L_{AC}$, the leakage current is 0.1 μA/mm at 90% of the *BV*, the $R_{ON,SP}$ is 25.8 mΩ·cm$^2$, and a calculated P-FOM of 3.8 GW/cm$^2$ is achieved. The device performance is even closer to the theoretical GaN limit below the SiC limit, as shown in Fig.5. Sine the device performance is mainly depended on the nature of the GaN itself, our results indicate that GaN is capable of device fabrication and the application in the UHV field.

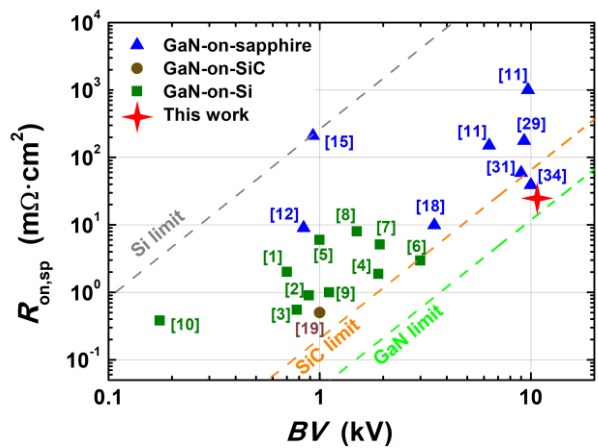

Fig.5 **The benchmark of the 10.6-kV SBD.** Our 10.6-kV SBD shows the highest *BV* and P-FOM, which is the closest to the theoretical limit of all GaN lateral SBDs so far.

**Demonstration of a practical large-size device with multi-finger structure**

The above experiments and analysis are based on the circular configuration for the principle study, and the device size is too small to be used in practical environment. In order to test the above results in the real environment, we fabricated large-size devices according to the most commonly used multi-finger structure in actual applications, as shown in Fig.6. The total size of the device is 5×5 mm².

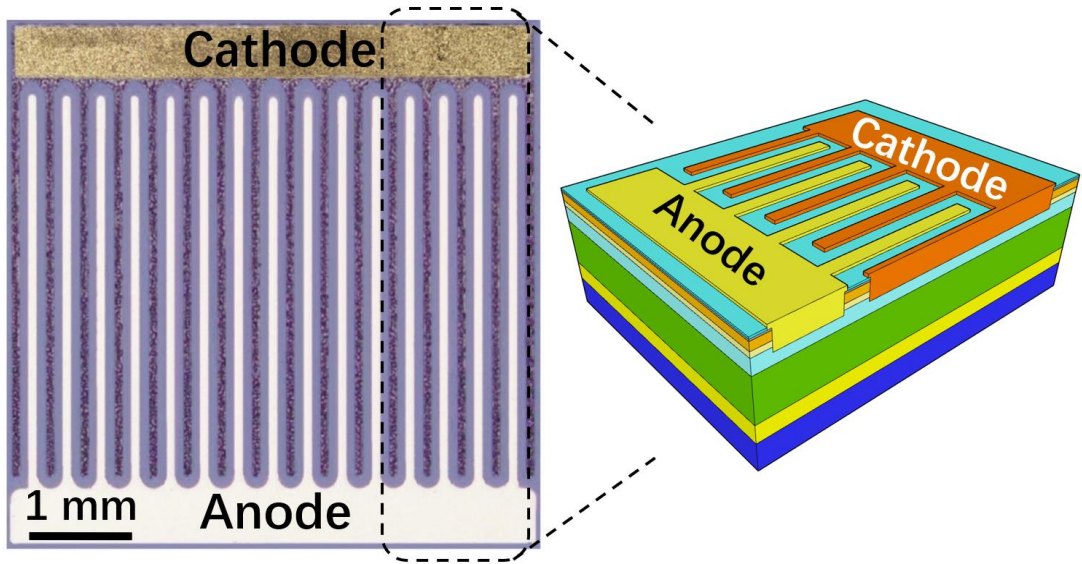

**Fig. 6 | Device structure. a,** photograph of top view of the fabricated multi-finger SBD with a $L_{AC}$ of 85 μm. **b,** Schematic bird view of the AlGaN/GaN multi-finger SBD on sapphire. The width of one anode finger ($W_A$) is 85 μm, the width of one cathode finger ($W_C$) is 85 μm, and the total size of the device is 5×5 mm².

Fig.7a shows the measured forward and reverse *I-V* characteristics of the 5×5 mm² multi-finger SBD. The forward current can reach 3.5 A at 3 V, which gives a $R_{ON,SP}$ of 61.6 mΩ·cm². The $R_{ON,SP}$ is a little larger than the one obtained in the small circular device, which could be caused by the fluctuation in the material quality and processing conditions in larger area. Fig.7b shows the reverse *I-V* characteristics, and gives a *BV* of 9.1 kV, which also is a little lower than the one obtained in the small circular device caused by the same reason. Thus, a calculated P-FOM of 1.34 GW/cm² is achieved in the multi-finger SBD. The decrease of the P-FOM compared to the small circular device is mainly caused by the self-heating under high current injection. In practical applications, high-power devices always have a good heat dissipation structure. Therefore, when a good heat dissipation structure is attached to our devices, it can be expected that the self-heating effect will be weakened. In general, our research results confirm that GaN can be used in the UHV field. Even if it is designed as a large-size device, the device still maintains most of the excellent performance.

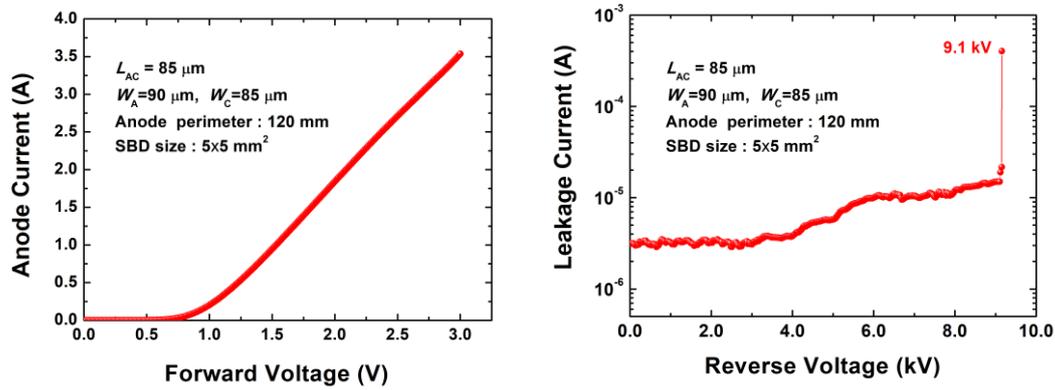

**Fig. 7 | Characteristics of the large-size SBD. a** Forward and **b** reverse I-V characteristics of GaN-on-sapphire lateral SBDs with multi-finger patterns.

## Conclusions

In summary, we have achieved 10.6-kV UHV AlGaN/GaN lateral SBDs on sapphire with single AlGaN/GaN channel and the cathode-to-anode spacing of 85 μm. Combining with a $R_{ON,SP}$ of 25.8 mΩ·cm$^2$ of the device, the P-FOM can be as high as 3.8 GW/cm$^2$. We also demonstrated a large size multi-finger SBD of 5×5 mm$^2$, the SBD exhibits a forward current of 3.5 A at 3 V, a *BV* of 9.1 kV and a high P-FOM of 1.34 GW/cm$^2$. These results provide clear and positive answers to the potential of GaN-based materials in UHV applications, and it is expected to realize the low-cost and high-performance applications of GaN materials in the field of UHV electronics.

## Method

**The fabrication of SBDs on sapphire.**

The SBD fabrication commenced with photolithography defining the SBD isolation area, then the $SiN_x$ was etched out by reactive ion etching (RIE) using the gas of $CF_4$, finally a 400 nm isolation mesa was etched out by inductively coupled plasma (ICP) using the mixture gas of $Cl_2/BCl_3$. In the second step, the anode recess area was also etched by ICP using the mixture of $Cl_2$ and $BCl_3$. Because the ICP etching usually causes the surface damages [42], so the sample was etched in a diluted KOH solution (0.1 mol/L) for 15 minutes in a water bath at 80 ℃ to remove the etch damages. The last step was evaporation of electrodes, we used Ti/Al/Ni/Au (30/150/30/100 nm) as the ohmic electrode followed by rapidly annealing at 850 ℃ for 30 s in $N_2$ and Pt/Au (50/300 nm) as anode.

**The fabrication of SBDs on Si.**

Firstly, the SBD mesa area was fabricated, commenced with growing a 250 nm $SiN_x$ passivation layer by plasma enhanced chemical vapor deposition (PECVD). Combined with the 50-nm in-situ $SiN_x$, the total thickness of $SiN_x$ is 300 nm. Then we use reactive ion etching (RIE) to etch out $SiN_x$ by the gas of $CF_4$, and we use ICP etching to form a 400-nm isolation mesa by using the mixture gas of $Cl_2/BCl_3$.

In the second step, the electrode recess area was defined by photolithography, and then the $SiN_x$ was etched out by the RIE. After that, the electrode recess area was etched by ICP using the mixture gas of $Cl_2$ and $BCl_3$ with a flow rate of 48/6 sccm. The power of the RF/ICP was selected as 100 W/300 W combined with a chamber pressure of 10 mTorr.

Next, the sample was etched in a diluted KOH solution for 15 minutes in a water bath at 80 ℃ to remove etch damages. The KOH treatment can effectively reduce the N deficiency at the surface caused by $Cl_2/Bl_3$ plasma, resulting in an improvement of the Schottky barrier height and the rectifying characteristics of SBDs [43,44].

Finally, the cathode and anode were prepared respectively, we chose Ti/Al/Ni/Au as the cathode metal and Pt/Au as the anode metal. The recessed ohmic contact allows the TiN contacting the 2-DEG channel directly after high temperature annealing, which can result in enhanced carrier tunneling and reduced contact resistance [45].

**Electrical measurements.**

The *I-V* characteristics were measured using a Keithley 4200. The reverse blocking voltage was measured using IWATSUCS12105C semiconductor curve tracer.

**References**

1. Lee, H. S. et al, 0.34 $V_T$ AlGaN/GaN-on-Si large Schottky barrier diode with recessed dual anode metal, *IEEE Electron Device Lett*., **36**, 1132–1134 (2015).

2. Tsou, C. W. et al, 2.07-kV AlGaN/GaN Schottky barrier diodes on silicon with high Baliga's figure-of-merit, *IEEE Electron Device Lett.*, **37**, 70–73 (2016).

3. Gao, J. N. et al, Low on-resistance GaN Schottky barrier diode with high $V_{ON}$ uniformity using LPCVD $Si_3N_4$ compatible self-terminated, low damage anode recess technology, *IEEE Electron Device Lett*., **39**, 859–863 (2018).

4. Zhang, T. et al, A 1.9-kV/2.61-mΩ·$cm^2$ Lateral GaN Schottky Barrier Diode on Silicon Substrate With Tungsten Anode and Low Turn-ON Voltage of 0.35 V, *IEEE Electron Device Lett.*, **39**, 1548–1551 (2018).

5. Zhang, T. et al, A > 3 kV/2.94 mΩ· $cm^2$ and Low Leakage Current With Low Turn-On Voltage Lateral GaN Schottky Barrier Diode on Silicon Substrate With Anode Engineering Technique, *IEEE Electron Device Lett.* **40**, 1583-1586 (2019).

6. Zhu, M. et al, 1.9-kV AlGaN/GaN lateral Schottky barrier diodes on silicon, *IEEE Electron Device Lett*., **36**, 375–377 (2015).

7. Lian, Y. W. et al, AlGaN/GaN Schottky barrier diodes on silicon substrates with selective Si diffusion for low onset voltage and high reverse blocking, *IEEE Electron Device Lett*., **34**, 981–983 (2013).

8. Lenci, S. et al, Au-free AlGaN/GaN power diode on 8-in Si substrate with gated edge termination, *IEEE Electron Device Lett.*, **34**, 1035–1037 (2013).

9. Lee, J. G. et al, Low turn-on voltage AlGaN/GaN-on-Si rectifier with gated ohmic anode, *IEEE Electron Device Lett*., **34**, 214–216 (2013).

10. Matioli, E. et al, Ultralow Leakage Current AlGaN/GaN Schottky Diodes With 3-D Anode Structure, *IEEE Transactions on Electron Devices.*, **60**, 3365-3370 (2013).

11. Zhang, A. P. et al, Lateral $Al_xGa_{1-x}N$ power rectifiers with 9.7 kV reverse breakdown voltage, *Appl. Phys. Lett.*, **78**, 823–825 (2001).

12. Park, K. et al, 1 kV AlGaN/GaN power SBDs with reduced on resistances, in *Proc. IEEE 23th Int. Symp. Power Semiconductor Devices IC's*, 223–22610 (2011), doi: 1109/ISPSD.2011.5890831.

13. Lei, J. C. et al, 650-V double-channel lateral Schottky barrier diode with dual-recess gated anode, *IEEE Electron Device Lett.*, **39**, 260–263 (2017).

14. Kamada, A. et al, High-voltage AlGaN/GaN Schottky barrier diodes on Si substrate with low-temperature GaN cap layer for edge termination, in *Proc. IEEE 20th Int. Symp. Power Semiconductor Devices IC's*, 225–228 (2008), doi: 10.1109/ISPSD.2008.4538939.

15. Lee, S. C. et al, High Breakdown Voltage GaN Schottky Barrier Diode employing Floating Metal Rings on AlGaN/GaN Hetero-junction, in *Proc. IEEE 17th Int. Symp. Power Semiconductor Devices & IC's*, 247-250 (2005), doi: 10.1109/ISPSD.2005.1487997.

16. Zhou, Q. et al, High reverse blocking and low onset voltage AlGaN/GaN-on-Si lateral power diode with MIS-gated hybrid anode, *IEEE Electron Device Lett.*, **36**, 660–662 (2015).

17. Hsin, Y. M. et al, A 600 V AlGaN/GaN Schottky barrier diode on silicon substrate with fast reverse recovery time, *Phys. Status Solidi C*, **9**, 949–952 (2012).

18. Lee, G. Y. et al, High-Performance AlGaN/GaN Schottky Diodes With an AlGaN/AlN Buffer Layer, *IEEE Electron Device Lett.*, **32**, 1519–1521 (2011).

19. Treidel, E. B. et al, Fast-switching GaN-based lateral power Schottky barrier diodes with low onset voltage and strong reverse blocking, *IEEE Electron Device Lett.*, **33**, 357–359 (2012).

20. Ma, J. et al, Multi-Channel Tri-Gate GaN Power Schottky Diodes With Low ON-Resistance, *IEEE Electron Device Lett.*, **40**, 275–278 (2019).

21. Lee, J. H. et al, AlGaN/GaN-Based Lateral-Type Schottky Barrier Diode With Very Low Reverse Recovery Charge at High Temperature, *IEEE Transactions on Electron Devices.*, **60**, 3032-3039 (2013).

22. Park, Y. et al, Low onset voltage of GaN on Si Schottky barrier diode using various recess depths, *IEEE Electronics Letters*, **50**, 1165-1167 (2014).

23. Lee, G. Y. et al, High-Performance AlGaN/GaN Schottky Diodes With an AlGaN/AlN Buffer Layer, *IEEE Electron Device Lett.*, **32**, 1519–1521 (2011).

24. Longobardi, G. et al, On the vertical leakage of GaN-on-Si lateral transistors and the effect of emission and trap-to-trap-tunneling through the AlN/Si barrier, *29th Int. Symp. Power Semiconductor Devices IC's*, 227–230 (2017), doi: 10.23919/ISPSD.2017.7988918

25. Zhou, C. H. et al, Vertical Leakage/Breakdown Mechanisms in AlGaN/GaN-on-Si Devices, *IEEE Electron Device Lett.*, **32**, 1132–1134 (2012).

26. Ozbek, A. M. and Baliga, B. J., Planar Nearly Ideal Edge-Termination Technique for GaN Devices, *IEEE Electron Device Lett.*, **32**, 300–302 (2011)

27. Vetury, R. et al, Direct Measurement of Gate Depletion in High Breakdown (405V) AlGaN/GaN Heterostructure Field Effect Transistors, *International Electron Devices Meeting 1998. Technical Digest,* 55-58 (1998), doi: 10.1109/IEDM.1998.746245.

28. Dora, Y. et al, High Breakdown Voltage Achieved on AlGaN/GaN HEMTs with Integrated Slant Field Plates, *IEEE Electron Device Letters*, **27**, 713-715 (2006).

29. Ishida, H. et al, GaN-based Natural Super Junction Diodes with Multi-channel Structures, *2008 IEEE International Electron Devices Meeting,* , 1-4, (2008) doi: 10.1109/IEDM.2008.4796636

30. Colón, A. et al, Demonstration of a 9 kV reverse breakdown and 59 mΩ·cm$^2$ specific on-resistance AlGaN/GaN Schottky barrier diode, *Solid-State Electron.*, **151**, 47–51 (2019).

31. Jiang, Y. et al, 10kV SiC MPS Diodes for High Temperature Applications, *Proceedings of the 2016 28th International Symposium on Power Semiconductor Devices and ICs (ISPSD)*, 43-46 (2016)

32. Ji, S. et al, Temperature-Dependent Characterization, Modeling and Switching Speed Limitation Analysis of Third Generation 10 kV SiC MOSFET, *IEEE Transactions on Power Electronics*, **33**, 4317-4327 (2018)

33. Xiao, M. et al, 10 kV, 39 mΩ·cm$^2$ Multi-Channel AlGaN/GaN Schottky Barrier Diodes, *IEEE Electronics Letters*, **42**, 808-811 (2021).

34. Xu, R., Chen, P., et.al. 2.5-kV AlGaN/GaN Schottky Barrier Diode on Silicon Substrate with Recessed-anode Structure, *https://arxiv.org/abs/2007.03163,* arXiv:2007.03163 (2020)

35. Xu, R., Chen, P., et.al., 2.7-kV AlGaN/GaN Schottky barrier diode on silicon substrate with recessed-anode structure, *Solid State Electronics*, **175**, 107953 (2021).

36. Xu, R., Chen, P., et.al., 3.4-kV AlGaN/GaN Schottky barrier diode on silicon substrate with engineered anode structure, *IEEE electron device letters,* **42**, 208-211 (2021).

37. Xu, R., Chen, P., et.al., 1.4-kV Quasi-Vertical GaN Schottky Barrier Diode With Reverse p-n Junction Termination, *IEEE Journal of the Electron Devices Society*, **8**, 316-320 (2020.

38. Würfl, J. et al, Device Breakdown and Dynamic effects in GaN Power Switching Devices: Dependencies on Material Properties and Device Design, *ECS Transactions*, **50**, 211-222 (2012).

39. Uren, M. J. et al, Buffer transport mechanisms in intentionally carbon doped GaN heterojunction field effect transistors, *Appl. Phys. Lett*, **104**, 263505-1-263505-5 (2014).

40. Kato, S. et al, C-doped GaN buffer layers with high breakdown voltages for high power operation AlGaN/GaN HFETs on 4-in Si substrates by MOVPE, *Journal of Crystal Growth*, **298**, 831-834 (2007).

41. Lu, B. et al, High Breakdown (> 1500 V) AlGaN/GaN HEMTs by Substrate-Transfer Technology, *IEEE Electronics Letters*, **31**, 951-953 (2010).

42. Yu, Z.G, Chen, P. et al, Influence of Dry Etching Damage on the Internal Quantum Efficiency of Nanorod InGaN/GaN Multiple Quantum Wells, *Chinese Physics Letters*, **29,** 078501-1-078501-4, (2012).

43. Choi, H. W. et al, Plasma-induced damage to n-type GaN ", *Appl. Phys. Lett*., **77**, 1795-1797, (2000).

44. Rickert, K. A. et al, n-GaNn-GaN surface treatments for metal contacts studied via x-ray photoemission spectroscopy, *Appl. Phys. Lett.,* **80**, 204-206 (2002).

45. Kamada, A. et al, High-voltage AlGaN/GaN Schottky barrier diodes on Si substrate with low-temperature GaN cap layer for edge termination, *Proc. IEEE 20th Int. Symp. Power Semiconductor Devices IC's,* 225–228 (2008) doi: 10.1109/ISPSD.2008.4538939.

**Acknowledgements**

The authors acknowledge Corenergy. Inc. and Enkris Semiconductor Inc. for the epitaxy of the wafers. This work is supported by National High-Tech R@D Project (2015AA033305), Jiangsu Provincial Key R&D Program (BK2015111), Collaborative Innovation Center of Solid State Lighting and Energy-saving Electronics, the Research and Development Funds from State Grid Shandong Electric Power Company and Electric Power Research Institute.

**Author contributions**

P.C. conceived the idea and designed the experiments. R.X. performed the experiments and numerical simulations and theoretical derivations. P.C. and R.X analyzed the results. T.Z. and K.C. contributed to the epitaxial growth. R.Z. and Y.Z. conducted research supervision. All the authors contributed to the manuscript writing.

**Competing interests**

The authors declare no competing interests.